\title{Stochastic Chemical Reactions in Micro-domains}

\documentclass[12pt]{article}
\usepackage{epsfig, graphicx,subfigure}
\usepackage{amsmath}
\usepackage{psfig}
\usepackage{latexsym}
\usepackage{amstext}

\newcommand{\mb}[1]{ \mbox{\boldmath$#1$} }
\newcommand{\ds}{\displaystyle}
\newcommand{\beq}{\begin{eqnarray}}
\newcommand{\eeq}{\end{eqnarray}}
\newcommand{\beqq}{\begin{eqnarray*}}
\newcommand{\eeqq}{\end{eqnarray*}}
\newcommand{\p}{\partial}

\newcommand{\x}{\mbox{\boldmath$x$}}
\newcommand{\n}{\mbox{\boldmath$n$}}
\newcommand{\J}{\mbox{\boldmath$J$}}
\newcommand{\y}{\mbox{\boldmath$y$}}

\font\bb=msbm10 at 12pt
\def\rR{\hbox{\bb R}}

\begin{document}
\pagestyle{plain}
\begin{center}
{\large \textbf{Stochastic Chemical Reactions in Micro-domains}}\\[5mm]
D. Holcman
\footnote{ Department of Mathematics, Weizmann Institute of Science, Rehovot
76100, Israel. D.H is incumbent to the Madeleine Haas Russell Career Development Chair.} \footnote{Keck-Center for Theoretical Neurobiology, Department of Physiology, UCSF 513 Parnassus Ave, San Francisco CA 94143-0444, USA.} and Z. Schuss\footnote{%
Department of Mathematics, Tel-Aviv University, Tel-Aviv 69978,
Israel}
\end{center}
\date{}

\begin{abstract}
\noindent  Traditional chemical kinetics may be inappropriate to
describe chemical reactions in micro-domains involving only a
small number of substrate and reactant molecules. Starting with
the stochastic dynamics of the molecules, we derive a
master-diffusion equation for the joint probability density of a
mobile reactant and the number of bound substrate in a confined
domain. We use the equation to calculate the fluctuations in the
number of bound substrate molecules as a function of initial
reactant distribution. A second model is presented based on a
Markov description of the binding and unbinding and on the mean
first passage time of a molecule to a small portion of the
boundary. These models can be used for the description of noise
due to gating of ionic channels by random binding and
unbinding of ligands in biological sensor cells, such as olfactory
cilia, photo-receptors, hair cells in the cochlea.
\end{abstract}

\section{Introduction}
Biological  micro-structures such as synapses, dendritic spines,
{ subcellular domains}, sensor cells and many other structures,
are regulated by chemical reactions that involve only a
small number of molecules, that is, between a few and up to
thousands of molecules. A chemical reaction that involves only 10
to 100 proteins can cause a qualitative transition in the
physiological behavior of a { given} part of a cell. Large
fluctuations should be expected in a reaction if so few molecules
are involved, both in transient and persistent binding and
unbinding reactions. In the latter case large fluctuations in the
number { of} bound molecules should force the physiological
state to change all the time, unless there is a specific mechanism
that prevents the switch { and} stabilizes the physiological
state. Therefore, a theory of chemical kinetics of such reactions
is needed to predict the threshold at which switches occur and to
explain how the physiological function is regulated in molecular
terms at a sub-cellular level.

A physiological threshold can be determined in molecular terms,
for example, when the number of activated molecules exceeds a
certain value. The standard theory of chemical kinetics is
insufficient for the determination of the threshold value, because
it is based on the assumption that there is a sufficiently large
number of reactant molecules and it describes the time evolution
of only the average number of  molecules. 
The standard theory of reaction-diffusion describes chemical reactions in terms of concentrations so that
fluctuations due to a small number of molecules are lost.

For example, in dendritic spines of neurons a flow of calcium
entering through the NMDA channels can induce a cascade of
chemical reactions. As calcium ions diffuse they can bind, unbind,
{and} leave the spine { without} binding. But if enough
calcium binds specific molecules, such as calmodulin, then {
certain proteins become activated, such as CaMK-II}, which are
involved in regulating synaptic plasticity \cite{Lisman94}. Now if
{ sufficiently many of them are activated at about the same time} and
thus the threshold is exceeded, additional changes can be induced at the
synapse level, affecting the physiological properties of a neuron.
In particular, such changes may include a modification of the
biophysical properties of some receptors and/or increase the
number of channels { at a specific area of the synapse, called
the postsynaptic density}. It is unclear how many CAMK-II are
needed {for crossing the threshold}, but the range is
somewhere between 5 to 50. It is remarkable that { as few as
5-50} molecules can control the synaptic weight \cite{Lisman94}.
The number of phosphorylated CAMK-II, activated after {a}
transient calcium flow, depends on the location of the proteins,
the location of the channels, the geometrical restrictions imposed
by the spine shape, and the state of the proteins. All of these
factors play a crucial role in determining the threshold.

The photoreceptor cells are another example, where fluctuations in
the number of bound molecules determine the physiological
limitations of the cell. Indeed, in the outer segment of cones and
rods of the retina, the total number of open channels fluctuates
continuously due to the binding and unbinding of specific gating
molecules to their receptors. These fluctuations are directly
converted into fluctuations of the membrane potential, which are
called ``dark noise'', and thus determine the signal to noise
ratio \cite{Baylor} for a photon detection. The fluctuation in the
number of open channels is regulated by the number of gated
molecules and depends on the geometry of the outer segment and the
distribution of channels. It is not clear what are the details of the biochemical
processes involved in regulating the number of open channels, but
interestingly, the signal due to a single photon event is
sufficient to overcome the noise amplitude in rods, but not in
cones, although their biochemical properties are similar. { As
the binding molecules diffuse in the cell they can bind and unbind
to channels, thus causing fluctuations in the number of open
channels. The fluctuation depends on the arrival time of the
binding molecules, called cGMP, to the channel binding sites}.

{In the mathematical description of the binding and unbinding
reactions, we model the diffusion of the particles as Brownian
motion and binding occurs when a particle reaches a binding site.}
The { binding} probability depends on the geometry of the
domain and on the distribution of the channels. A channel opens
when it binds to two or three cGMP molecules and in the absence of
light, the number of open channels is small, approximatively 6 to
10 per micro-domain in a mammalian rod, when there are only 60
cGMP molecules. Due to the random binding and unbinding of the
molecules to the channels the number of open channels is a
stochastic process.

In this paper we start with the stochastic dynamics of the
reactant molecules in a micro-domain and derive a master-diffusion
equation for the joint probability density of the mobile reactant
and the number of bound immobile substrate { molecules}. We use
the equation to calculate the fluctuations in the number of bound
substrate molecules as a function of initial reactant
concentration. We apply the present theory to the computation of
the mean and variance of the fluctuation in the number of open
channels, and find their dependence on the initial number of the
mobile reactant, the geometry of the { compartment}, and the
distribution of channels.

Our model can predict {the} fluctuation { intensity} produced by
binding and unbinding to channels in confined compartments, such
as the a compartment (the space between two consecutive disks) of
cone and rod outer segment, or any other sub compartment of a
sensor cell. Such a prediction can clarify part of the noise
generation. At the present time the noise in a confined
micro-domain cannot be directly measured. Instead, excised patch
measurements are done \cite{Yau,Koko}, in which the cell structure is destroyed.
Thus computation and simulation of mathematical models are the
only tools for studying noise in { this} biological context.

\section{A Stochastic model of a non-Arrhenius chemical reaction}\label{firstM}

\subsection{Chemical reaction}
We consider two species of reactants, the mobile reactant $M$ that diffuses
in a bounded domain $\Omega $, and the stationary
substrate $S$ (e.g., a protein). The boundary $\partial \Omega $ of the
domain $\Omega $ is partitioned into an absorbing part $\partial \Omega _{a}$
(e.g., pumps, exchangers, another substrate that forms permanent bonds with $%
M $, and so on) and a reflecting part $\partial \Omega _{r}$ (e.g., a cell
membrane). In this model the volume of $M$ is neglected. We assume that
there are binding sites on the substrate. In terms of traditional chemical
kinetics the binding of $M$ to $S$ follows the law
\begin{eqnarray}  \label{CRE}
M+S_{free}
\begin{array}{l}
k_{f} \\
\rightleftharpoons \\
k_{b}%
\end{array}
MS,  \label{reaction}
\end{eqnarray}
where $k_{f}$ is the forward binding rate constant, $k_{b}$ is the backward
binding rate constant, and $S_{free}$ is the unbound substrate.

However, when only a small number of reactant and substrate
molecules are involved in the reaction, as is the case in a
micro-domain in a biological cell, this reaction has to be
described by a molecular model, rather than by concentrations. The
description of this reaction on the molecular level begins with
the following definitions:

\begin{itemize}
\item $M(t)$ = number of unbound $M$ particles at
time $t$
\item $S_{\Delta}(\x,t)$ = number of free sites in volume $\Delta\x$ at
time $t$
\item $S(t)$ = number of unbound binding sites at time $t$
\item $MS(t)$ =$M(0)-M(t)$ = number of bound $M$
particles at time $t$.
\item $s_0(\x)$ = initial density of substrate\\
\item $S_{\mbox{max}}=\ds\int_{\Omega}s_0(\x)\,d\x$ = total number of
binding sites in $\Omega$.
\end{itemize}
The $M(0)$ reactant particles are initially distributed with
probability density $m_0(\x)$. The initial density $s_0(\x)$
integrates to $S(0)$. We assume that the $M$ particles diffuse and
we denote by $\x(t)$ the random trajectory of an unbound $M$
particle. We consider a small volume $\Delta \x$ about $\x$, that
contains initially $s_0(\x)\,\Delta\x$ free binding sites and
$m_0(\x)\,\Delta\x$ unbound $M$ particles.

The joint probability of an $M$ trajectory and the number of bound
sites in the volume $\Delta \x$  is
 \beq
p(\x,S,t\,|\,\y)\,\Delta\x=
\Pr\left\{\x(t)\in\x+\Delta\x,\,S_{\Delta}(\x,t)=S\,|\,\x(0)=\y\right\}.
\label{jpdf}
 \eeq
The function $p(\x,S,t\,|\,\y)$ is the joint probability density
to find an $M$ particle and $S_{\Delta}(\x,t)$ free binding sites
at $\x$ at time $t$, conditioned by the initial position $\y$ of
the $M$ particle.

\begin{figure}
\centering \psfig{figure = 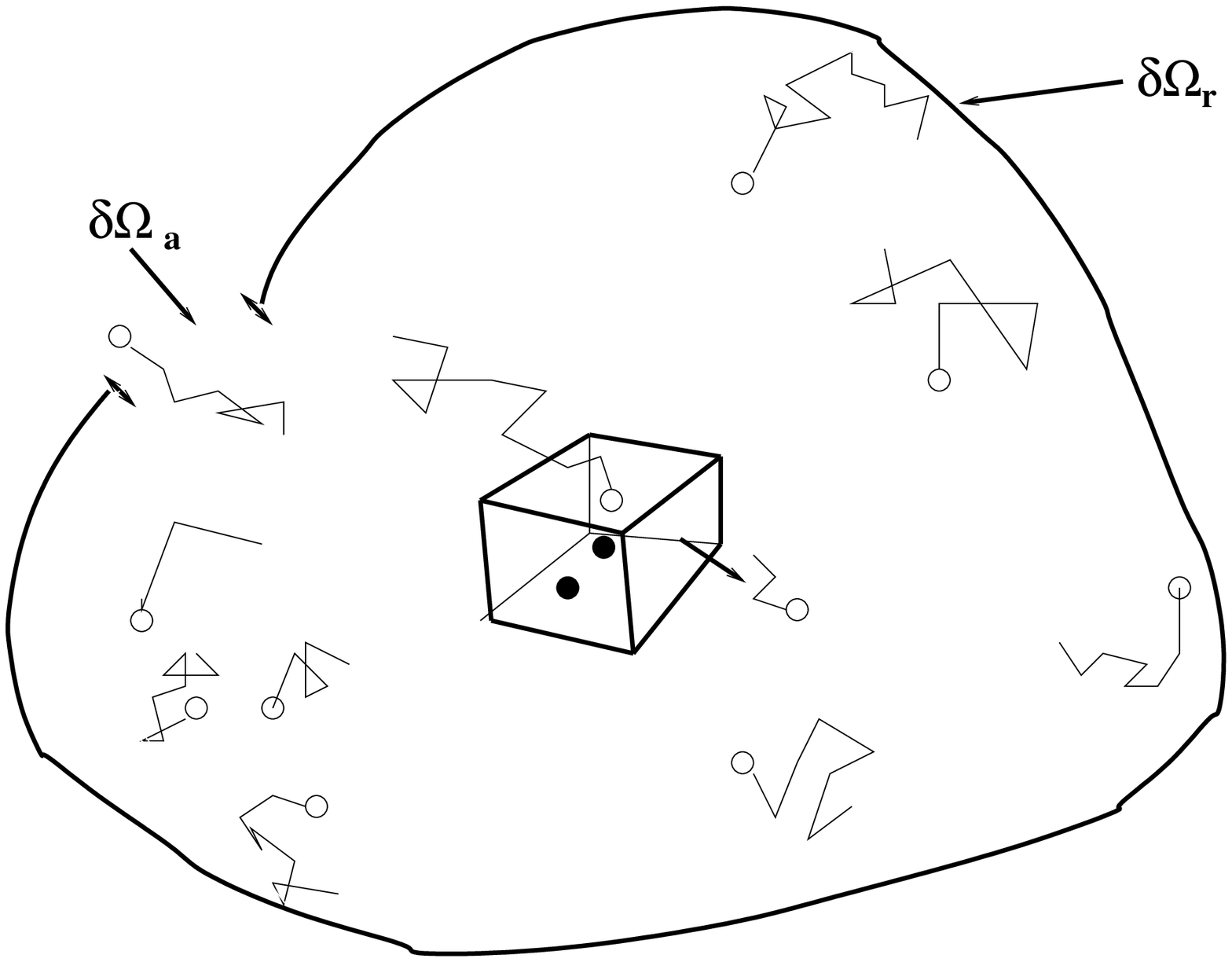,height =105mm,width=105mm} 
\caption{\small {{\bf Chemical reaction inside a microdomain}. Each free particle (white circle) moves according to a Brownian motion until it binds to a free site (black color). An occupied site cannot accept  any other particles. A molecule is composed of a finite number of sites. Parts of the microdomain is  reflective and the rest is absorbing.}}
\label{inside}
\end{figure}

The marginal probability density of an $M$ trajectory is
\begin{eqnarray*}
p(\x,t\,|\,\y) = \sum_{\mbox{free $S$}} p(\x,S,t\,|\,\y),
\end{eqnarray*}
where the sum is over all free binding sites in the volume $\Delta
\x$ . The number of free $M$ molecules in the volume $\Delta \x$
at $\x$ is
\begin{eqnarray*}
M_{\mbox{free}}(\x,t) =M_0\,\Delta\x\,\sum_{\mbox{free
$S$}}\int_\Omega p(\x,S,t|\y)m_0(\y)\,d\y.
\end{eqnarray*}
The joint probability density function of $\x$ and $S$ is
 \[p(\x,S,t)=\int_\Omega p(\x,S,t|\y)m_0(\y)\,d\y.\]
The two-dimensional process
$\left(M_{\mbox{free}}(\x,t),S(\x,t)\right)$ is Markovian. The
evolution of $p(\x,S,t)$, is governed by the diffusion of $M$
particles in and out of the volume $\Delta \x$ , and their binding
and unbinding inside the volume $\Delta \x$ . The influx in the
time interval $[t,t+\Delta t]$ is
 \begin{eqnarray*}
&&\oint_{\ds\p\mbox{Volume $\Delta
\x$}}\mb{J}(\x,S,t\,|\,\y)\cdot\mb{n}(\x) \,dS_{\x}\Delta
t=\\
&&\\
&&\int_{\mbox{Volume $\Delta \x$}}[p(\x,S,t+\Delta
t\,|\,\y)-p(\x,S,t\,|\,\y)]\,d\x,
\end{eqnarray*}
which represents diffusion with coefficient $D$. Additional change
in the contents of the volume $\Delta \x$  is due to the binding
and unbinding of $M$ particles to the substrate. When there are
$S$ free binding sites in the volume $\Delta \x$ (see Figure \ref{inside}), the probability
that one $M$ particle binds to a free site in the volume $\Delta
\x$  { in the time interval} $[t,t+\Delta t]$ is proportional
both to $S$ and to the number $M_{\mbox{free}}(\x,t)$ of free $M$
particles in the volume $\Delta \x$ . It is given by
 \begin{eqnarray*}
k_1M_0p(\x,S,t)S \Delta \x \, \Delta t.
\end{eqnarray*}
The probability that one particle unbinds in the volume $\Delta
\x$  in this time interval is proportional to the number
$s_0(\x)\Delta\x-S$ of bound sites in the volume $\Delta \x$ ,
given by
 \[k_{-1}[s_0\Delta\x-S] \Delta t.\]
Thus the probability of $S$ free sites when no change occurred in
the number of free sites is
 \begin{eqnarray*}
 p(\x,S,t) \Delta \x\left\{1-k_1M_0 p(\x,S,t)
 S\,\Delta \x\,\Delta t-
 k_{-1}[s_0(\x)\Delta\x-S] \Delta t\right\} .
 \end{eqnarray*}
The number of free sites can change to $S$ at the end of the
interval $[t,t+\Delta t]$ if it was $S+1$ at the beginning and one
bond was formed, or if it was $S-1$ and one particle was unbound.
The probability of this event, as calculated above, is
 \begin{eqnarray*}
k_1M_0\,\Delta \x\, (S+1) p(\x,S+1,t)\,\Delta
t+k_{-1}[s_0(\x)\Delta\x-S+1)]\Delta t.
\end{eqnarray*}
The probability of $\x,S$ at time $t+\Delta t$ is therefore
\begin{eqnarray*}
&&p(\x,S,t+\Delta t) \Delta \x=\\
&&\\
 &&-\nabla\cdot\mb{J}(\x,S,t\,|\,\y)\,\Delta\x  +\\
 && \\
 &&p(\x,S,t) \Delta \x\{1-M_0k_1\Delta \x\,S p(\x,S,t)\Delta
 t-k_{-1}[s_0(\x)\Delta\x-S] \Delta t\}+\\
&&\\
&& M_0k_1(S+1)  p^2(\x,S+1,t)\,(\Delta \x)^2\Delta t+\\
&&\\
&&  k_{-1}[s_0(\x)\Delta\x-S+1]p(\x,S-1,t) \Delta \x\Delta t.
 \end{eqnarray*}
{ for $S=0,1,2,\dots,S_{\mbox{max}}$, the coupled partial
differential equations}
 \beq
\frac{\p p(\x,S,t) }{\p t} &=& -\nabla\cdot\mb{J}(\x,S,t)
 - K_1 p^2(\x,S,t)S -k_{-1}[S_0(\x)-S]p(\x,S,t) \nonumber\\
&&\label{eqff}\\
&+& K_1
(S+1)p^2(\x,S+1,t)+k_{-1}[S_0(\x)-S+1]p(\x,S-1,t)\nonumber,
 \eeq
where by definition $\mb{J}(\x,S,t)$ is the joint probability flux
at position $\x$ at time $t$, and $S$ proteins are free. It is
defined in the diffusion case by
 \beq \mb{J}(\x,S,t) = -D \nabla
p(\x,S,t).
 \eeq
{ The} new (forward) binding rate is
\begin{eqnarray*}
K_1 =M_0 k_1 \Delta\x
\end{eqnarray*}
and
 \[S_0(\x)=s_0(\x)\,\Delta\x=\mbox{total number
of binding sites in the volume $\Delta \x$ }.\] Thus $K_1$ is the
probability flux into the binding sites. The boundary conditions
on $S$ are
 \beq
 p(\x,S,t)=0\quad\mbox{for $S<0$ and $S>S_0(\x)$}.\label{SBC}
 \eeq

 \vspace{3mm}

{\bf \noindent Remark 1.} By summing equation (\ref{eqff}) over
$0\leq S\leq S_0(\x)$ and using the boundary conditions
(\ref{SBC}), we obtain
 \beq\label{eqffM}
  \frac{\p M_{\mbox{total}}(\x,t) }{\p t} = -\nabla\cdot\mb{J}_M(\x,t),
 \eeq
for the marginal density of the $M$ particles. It means { that}
all bound and free $M$ particles effectively diffuse.\\

{\bf \noindent Remark 2.} When at specific locations there can be
at most one binding site, the system (\ref{eqff}) reduces to the
coupled equations
 \beq\label{eqff2}
\frac{\p p(\x,0,t) }{\p t} =
D\Delta p(\x,0,t) -k_{-1}p(\x,0,t) + K_1 p^2(\x,1,t)\nonumber  \\
\\
\frac{\p p(\x,1,t) }{\p t} = D\Delta p(\x,1,t) +k_{-1}p(\x,0,t)
-K_1 p^2(\x,1,t).\nonumber
 \eeq
Here $S_0(\x)$ can take the values 0 or 1. When no molecules can
escape from a bounded domain, the flux associated with $p(\x,S,t)$
satisfies the reflective boundary condition
 \beq
 \left.\mb{J}\cdot \n\right|_{\p \Omega} =0.\label{reflect}
 \eeq

The initial condition, when no substrate is bound, is given by
$p(\x,0,0) =m_{0}(\x)$, hence $p(\x,1,0) =0$. When the total
number of $M$ particles stays constant (i.e., no particles leave
the domain), adding equations (\ref{eqff2}) gives in the steady
state
 \beq
 p(\x,0)+p(\x,1)=\frac1{|\Omega|}.\label{conserve}
 \eeq

If the $M$ particles can escape the compartment, e.g., by being
absorbed in a part of the boundary $\p\Omega_a$, the condition
(\ref{reflect}) should be changed to
 \beq
 \left.\mb{J}\cdot \n\right|_{\p\Omega-\p \Omega_a} =0\label{reflecta}
 \eeq
and
 \beq
 \left.p(\x,S,t)\right|_{\p\Omega_a}=0.\label{absorba}
 \eeq
In this case (\ref{conserve}) no longer holds.\\

\noindent {\bf Remark 3:} Obviously, $S_0(\x)$ takes only integer
values. We assume that its discontinuities are located on smooth
interfaces. The density $p(\x,S,t)$ and the normal component of
the flux $\mb{J}(\x,S,t)\cdot\mb{n}(\x)$ are continuous across the
interfaces for all $S=0,1,\dots,S_{\mbox{max}}$.

\subsection{Moments of the pdf}
Statistical moments of the pdf can be computed from  equation
(\ref{eqff}). The average and the standard deviation of the number
of bound proteins are evaluated for equation (\ref{eqff2}). The
mean of the number of  bound proteins at time $t$ is given by
 \beq
\langle S_b(t)\rangle = \int_{\Omega} S_0(\x)p(\x,1,t) d\x.
 \eeq
The standard deviation is given by
 \beq \label{sd}
\sigma^2(t) = \langle S_b^2(t)\rangle-\langle S_b(t)\rangle^2  =
\int_{\Omega} S^2_0(\x)p(\x,1,t)\, d\x - \left(\int_{\Omega}
S_0(\x)p(\x,1,t)\, d\x\right)^2.
 \eeq
When the proteins are uniformly distributed on a subset $\Omega'
\subset \Omega$, the distribution $S_0(\x)$ is given by the
characteristic function of the subset $\Omega'$,
 \beq
S_0(\x) = \chi_{\Omega'}(\x)\frac{N_0}{\mid \Omega' \mid},
 \eeq
where the total number of binding sites is
 \[N_0=\int_{\Omega} S_0(\x)\, d\x.\]
We obtain the standard deviation of bound sites from the
expression (\ref{sd}) as
 \beq\label{sigma2}
  \sigma^2(t) &=&
\langle S_b^2(t)\rangle-\langle S_b(t)\rangle^2\\
&&\nonumber\\
&=& \left(\frac{N_0}{\mid \Omega' \mid}\right)^{2}\left(
\bar{S_0}(t)-\bar{S_0}^2(t)\right)= \left(\frac{N_0}{\mid \Omega'
\mid}\right)^{2} \bar{S_0}(t)(1-\bar{S_0}(t)),\nonumber
 \eeq
where

 \beq \label{Pro}
\bar{S_0}(t) = \int_{\Omega'} p(\x,1,t)\, d\x
 \eeq
is the fraction of bound sites. Note that
 \[\bar{S_0}(t) \leq 1.\]

\subsection{Standard deviation of the number of bound protein in some cases}
We consider the one-dimensional case where $\Omega=[0,L]$ and
$S_0(x)$ is either 0 or 1 in intervals. In the steady state the
system (\ref{eqff2}) is
  \beq
0&=& D\Delta p(x,0) -k_{-1}S_0(x) p(x,0) + K_1 p^2(x,1)\nonumber  \\
&&\nonumber\\
0&=& D\Delta p(x,1) +k_{-1}S_0(x) p(x,0) -K_1 p^2(x,1)\label{eqff3}\\
&&\nonumber\\
\frac{1}{L} &=&  p(x,0)+ p(x,1), \nonumber
 \eeq
which reduces to
 \beq\label{eqq4}
 D p''(x,1)
+k_{-1}S_0(x)\left[\frac{1}{L}-p(x,1)\right] -K_1 p^2(x,1)=0.
 \eeq
We convert to densities by setting
 \[c_M(x,1)=M_0p(x,1).\]
Then (\ref{eqq4}) becomes
 \beq\label{eqq5}
 D c_M''(x,1)
+k_{-1}S_0(x)\left[\frac{M_0}{L}-c_M(x,1)\right]
-\frac{K_1}{M_0}c_M^2(x,1)=0.
 \eeq
The function $c_M(x,1)$ is supported where the protein are
located.

When the boundary conditions are reflective for the $M$ trajectories, using the uniqueness of the solution, then
at a point $x$, where $S_0(x)$ is supported is
 \beqq
c_M(x,1) =\frac{2k_{-1}S_0(x)M_0/L}{k_{-1}S_0(x)+ \sqrt{
(k_{-1}S_0(x))^2+4 K_1k_{-1}S_0(x)/L }}.
 \eeqq
In particular, If the substrate is uniformly distributed in intervals $S_0(x)=N_S/L$ and $M_0\ll N_S$, then the fraction of bound $M$ particles is
 \beq
 p_M=\frac1{M_0}\int_0^Lc_M(x,1)\,dx=\frac{2}{1+ \sqrt{
1+4\ds \frac{M_0k_1\Delta x}{N_Sk_{-1}}}}\sim 1,\label{bound}
 \eeq
which means that practically all $M$ particles are bound. If
$M_0\gg N_S$, then  (\ref{bound}) gives
 \beq
  p_M=\frac1{M_0}\int_0^Lc_M(x,1)\,dx\sim\sqrt{\frac{N_Sk_{-1}}{M_0k_1\,\Delta
  x}}\ll1.
  \eeq

In this case, the variance of the fluctuations in the number of
bound $M$ particles, which is the same as the number of bound
sites, as a function of $M_0$ and $N_S$, is given by
 \beq
 \sigma^2_S(M_0)=p_M(1-p_M)=\frac{2}{1+ \sqrt{
1+4\ds \frac{M_0k_1\Delta x}{N_Sk_{-1}}}}\left(1-\frac{2}{1+
\sqrt{ 1+4\ds \frac{M_0k_1\Delta
x}{N_Sk_{-1}}}}\right).\label{Var}
 \eeq
The graph of $\sigma^2_S(M_0)$ vanishes for both $M_0\to0$ and
$M_0\to\infty$ and has a unique finite maximum, as in figure \ref{variance1}


\subsection{General equations when binding proteins are located on the boundary of the domain}
\begin{figure}
\centering \psfig{figure = 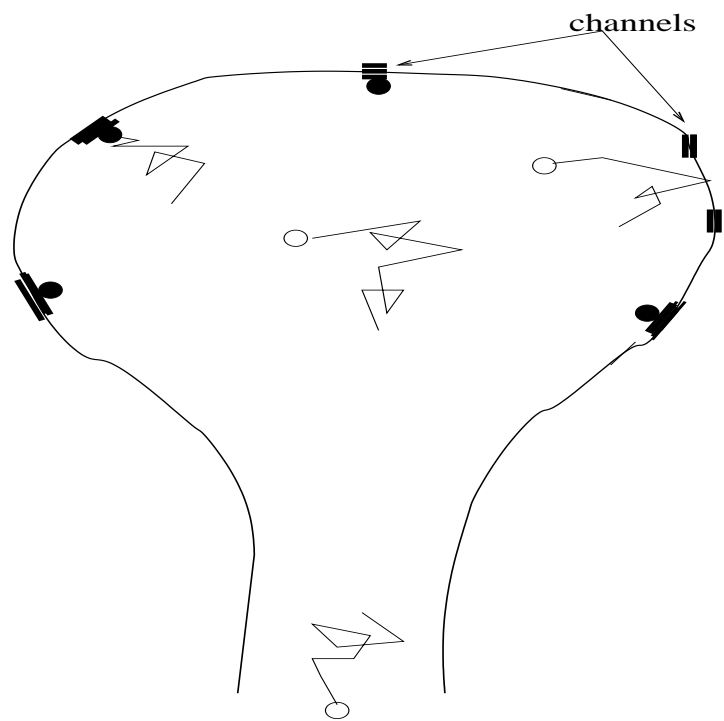, height =95mm,width=95mm}  
\caption{\small {{\bf Schematic reaction in a microdomains where channels are located on the boundary.} Molecules move according to a Brownian motion inside the domain. When a particle hits a free channel, it can bound and stays there for a certain amount of time. Particles can escape through the absorbing boundary. }}
\label{boundary}
\end{figure}

We consider the same problem, but with binding sites located on
the boundary $\p \Omega$ with surface density $S_0(\x)$ (see figure \ref{boundary}). This may
represent for example binding to gated channels, which open when
they bind agonist molecules. Some channels may need to bind
several agonist molecules to open, however, we consider here the
case that a single agonist molecules opens the channel upon
binding. We assume that the agonist molecule cannot escape
$\Omega$.

The initial $M_0$ agonist molecules diffuse in $\Omega$ and are
reflected at $\p\Omega$ at non binding sites, but can bind to a
free binding site on a protein channel in the membrane with a
certain forward binding rate $k_1$. When this occurs the channel
opens and stays open as long as the agonist is bound. The bound
agonist is released from the bound state at a backward rate
$k_{-1}$. Then the number of open channels is the number of
missing $M$ particles in $\Omega$. We write
 \beqq
 M_0=\int_\Omega
 c(\x,t)\,d\x+\int_{\p\Omega}S_{\mbox{bound}}(\x,t)\,dS_{\x.}
 \eeqq
The equations for the density of $M$ particles in $\Omega$ and in
$\p\Omega$ is
 \beq
\frac{\partial c(\x,t)}{\partial t}& = &D\Delta c(\x,t)\nonumber\\
&&\nonumber\\
-D\frac{\p
c(\x,t)}{\p\n}&=&\J(\x,t)\cdot\n(\x){\Bigg{|}}_{\x\in\p\Omega}=\dot S_{\mbox{bound}}(\x,t)\nonumber\\
&&\nonumber\\
&=& k_1
c(\x,t)[S_0(\x)-S_{\mbox{bound}}(\x,t)]-k_{-1}S_{\mbox{bound}}(\x,t).\label{dotsb}
 \eeq

The probability of $k$ bound sites on the boundary at time $t$
satisfies the equations
 \beq
\dot P_k(t)&=&-P_k(t)K_1\oint_{\p\Omega}c(\x,t)\left[S_0(\x)-
S_{\mbox{bound}}(\x,t)\right]\,dS_{\x}
-kP_k(t)k_{-1}\nonumber\\
&&\nonumber
 \\&&+P_{k-1}(t)K_1\oint_{\p\Omega}c(\x,t)\left[S_0(\x)
-S_{\mbox{bound}}(\x,t)\right]\,dS_{\x}
+(k+1)P_{k+1}(t)k_{-1},\label{Pkdot},
 \eeq
\beq
\dot P_0(t)&=&-P_0(t)K_1\oint_{\p\Omega}c(\x,t)\left[S_0(\x)-
S_{\mbox{bound}}(\x,t)\right]\,dS_{\x} +P_{1}(t)k_{-1},\label{Pkdot0},
 \eeq
 \beq
\dot P_{S_0}(t)&=&-P_{S_{0}}(t)K_1\oint_{\p\Omega}c(\x,t)\left[S_0(\x)-
S_{\mbox{bound}}(\x,t)\right]\,dS_{\x}
-S_0P_{S_0}(t)k_{-1}\nonumber\\
&&\nonumber
 \\&&+P_{S_0-1}(t)K_1\oint_{\p\Omega}c(\x,t)\left[S_0(\x)
-S_{\mbox{bound}}(\x,t)\right]\,dS_{\x}
 \eeq
where, according to eq.(\ref{dotsb})
 \beqq
 \dot S_{\mbox{bound}}(\x,t)= k_1
c(\x,t)[S_0(\x)-S_{\mbox{bound}}(\x,t)]-k_{-1}S_{\mbox{bound}}(\x,t).
 \eeqq
Here $0\leq k\leq S_0$, where
 \beqq
 S_0=\oint_{\p\Omega}S_0(\x)\,dS_{\x},
 \eeqq
that is, $P_{-1}(t)=P_{S_0+1}(t)=0$.

The moments of the number of bound sites are
 \beqq
 \langle S_b(t)\rangle=\sum_{k=1}^{S_0}kP_k(t),\quad \langle
 S_b^2(t)\rangle=\sum_{k=1}^{S_0}k^2P_k(t).
 \eeqq
The variance in the number of bound sites is
 \beqq
 \sigma^2(t)=\langle
 S_b^2(t)\rangle-\langle S_b(t)\rangle^2.
 \eeqq

\section{The fluctuation in the number of particles in a push-pull chemical reaction}
A push-pull chemical reaction consists of a source that produces particles at a given rate and a sink or a killing term that destroy or remove the particles from the system with its rate. When the movement of the particles is driven by diffusion and the source and the sink are not uniformly distributed, the number of particles fluctuates.  In this section, we propose an approach to estimate the mean number and the fluctuations. A permanent regime imposes that the rate of the sink and the source satisfy some specific conditions. At equilibrium, in the limit of a large number particles,  the
push-pull mechanism reaches a steady state and the number of particles does not fluctuate and can be computed using the rate constant. But for a small number, the analysis requires to study separately the dynamics of the particles and especially the law of injection by the source.

In the neurobiological context, many biochemical reactions are based on push-pull mechanisms: at a given location, a source produces molecules and somewhere else, an enzyme modeled as a sink, destroys the molecules with a certain efficassy. In general, sinks and sources are not uniformly distributed, which induces fluctuations.  As an example, gating molecules that can open channels are produced by such push-pull mechanism, and the fluctuations in the number of molecules induce a fluctuation in the number of open  channels, which is reveals at the cellular level. It is of particular interest to examine the situation  where molecules move by diffusion, the enzyme sink is represented by a single molecule and the sources are uniformly distributed. For example, this occurs inside a compartment of photo-receptors for the regulation of cGMP molecules.

\subsection{The push-pull mechanism}\label{ppm}
The sink of the push-pull mechanism is modelled by a killing term
located at a single point. The killing term is a Dirac function,
({ see \cite{HMS} for the exact mathematical interpretation}).
The source is assumed to be uniformly distributed and molecules
are produced at a constant rate $\gamma$.

We assume that the molecules are independent and their movement
{ in a domain $\Omega$ can be described by the stochastic
differential equations
 \beq \dot{\x}_k = \mb{b}(\x_k)
+\sqrt{2D}\, \dot{\mb{ w}}_k, \quad k=1,\ldots,N,
 \eeq
where $\mb{ b}(\x)$ is a drift vector}. Since the molecules cannot
escape, reflecting boundary conditions are imposed at the boundary
$\p\Omega$. If a particle is injected in $\Omega$ at time $t=0$,
its pdf $p(\x,t)$ satisfies the Fokker-Planck equation
 \beq \label{dynb}
 \frac{\p p(\x,t)}{\p t} &=&-\nabla\cdot{\J}(\x,t)-k_1\delta(\x-\x_1) p(\x,t)
\nonumber \\
& &\\
 \left.\J\cdot \n\right|_{\p\Omega}& =&0,\nonumber
 \eeq
where the probability flux { density vector} is given by
 \beq
 \J(\x,t)=-D\nabla p(\x,t)+\mb{b}(\x) p(\x,t).\label{pflux}
  \eeq
At any moment of time, when there are $N$ particles inside
$\Omega$, the concentration $c(\x,t)$ is given by
 \beq
 c(\x,t) = Np(\x,t).
  \eeq
Particles are injected with a Poisson stream at a rate $\gamma$,
{ that is, at independent and identically distributed
inter-injection times, whose pdf is}
 $$f(t) =\gamma e^{-\gamma t}.$$
{The mean inter-injection time is}  $E(T)=\gamma^{-1}$. The
probability that a molecule injected at time $t=0$ at a point $\y$
survives at time $t$ is
 \beq
 S_{\y}(t) =Pr\{ \x(t) \in \Omega \} = \int_\Omega p(\x,t\,|\,\y)\,d\x.
 \eeq
When $\y=\mb{0}$, we denote $S(t)=S_{\mb{0}}(t)$. To compute the
mean and the variance of the number of molecules $N(t)$ surviving
in $\Omega$ at time $t$, we use here the renewal equation
\cite{kleinrock,karlin,amit},
\begin{eqnarray}
\Pr\{N(t)=0 \} & = & \Pr\{ \x(t) \notin \Omega \}\cdot \left[
\int_0^
tf(s)\Pr\{N(t-s)=0\}\,ds + \int_t^\infty f(s)\,ds\right] \\
&&\nonumber\\
\Pr\{N(t)=1 \} & = & \Pr\{\x(t)\in \Omega \}\cdot \left[ \int_0^t
f(s)\Pr\{N(t-s)=0\}\,ds + \int_t^\infty f(s)\,ds \right] \nonumber \\
&&\nonumber\\ & & + \Pr\{\x(t) \notin \Omega\}\cdot \int_0^t
f(s)\Pr\{N(t-s)=1\}\,ds. \\
&&\nonumber\\ \Pr\{N(t)=n\} & = & \Pr\{\x(t)\in \Omega \}\cdot
\int_0^t
f(s)\Pr\{N(t-s)=n-1\}\,ds \nonumber \\
&&\nonumber\\ & & + \Pr\{\x(t) \notin \Omega\}\cdot \int_0^t
f(s)\Pr\{N(t-s)=n\}\,ds, \quad n > 1.
\end{eqnarray}
The expected number $EN(t)$ { of molecules surviving in
$\Omega$ at time $t$} is
\begin{eqnarray}
EN(t) & = & \sum_{n=1}^\infty n\Pr\{N(t)=n\} \nonumber \\
&&\nonumber\\
 & = & \Pr\{\x(t)\in \Omega \}\cdot \left( \int_0^t
f_L(s)EN(t-s)\,ds + \int_0^\infty f(s)\,ds \right) \nonumber
\\
&&\nonumber\\
 && + \Pr\{\x(t)\notin \Omega\}\cdot \int_0^t
f(s)EN(t-s) \,ds \nonumber \\
&&\nonumber\\ & = & \Pr\{\x(t)\in \Omega \} + \int_0^t
f(s)EN(t-s)\,ds.\label{ieq}
\end{eqnarray}
The integral equation (\ref{ieq}) is solved by the Laplace
transform as
\begin{eqnarray}
\bar n(\tau) = \frac{\bar{S}(\tau)}{1-\bar{f}(\tau)},
\end{eqnarray}
where $\bar{S}(\tau)$ is the Laplace transform of $\Pr\{\x(t)\in \Omega \}$ when the initial position of insertion is 0.
\begin{eqnarray}
\bar{f}(\tau) = \frac{\gamma}{\gamma +\tau}
\end{eqnarray}
and $\bar{S}(\tau)$ is the Laplace transform of the survival
probability. Therefore the Laplace transform of $EN(t)$ is given
by
\begin{eqnarray}
\bar n(\tau) = \frac{(\gamma+\tau) \bar{S}(\tau)}{\tau}.
\end{eqnarray}

\subsubsection{Computation of  $\bar{S}(\tau)$ in a driftless
one-dimensional model}

We consider the one-dimensional equation (\ref{dynb}) in
$\Omega=[0,L]$ and $\mb{b}(\x)=\mb{0}$. We assume, for simplicity,
that $L=\pi$ and $D=1$.

The survival probability given by $S(t) = \ds\int_0^L p(x,t\,|\,y)
dx$ is also (see \cite{HMS}) equal to\begin{eqnarray} S(t) =
1-k_1\int^t_0 p(x,t\,|\,y) dx.
\end{eqnarray}
The Laplace transform is
\begin{eqnarray}
\bar{S}(\tau) = \frac{1-k_1\bar{p}(x_1,\tau\,|\,y)}{\tau}.
\end{eqnarray}
$\bar{p}(x_1,\tau\,|\,y)$ can be computed using the Green's
function for the Neumann problem for (\ref{dynb}), given by
\begin{eqnarray}
G(x,t\,|\,y) =1+\frac{2}{\pi} \sum_{n=1}^\infty e^{-n^2t}\cos nx\cos ny .
\end{eqnarray}
Following the computations of \cite{HMS}, an integral representation of the solution
 of equation \ref{dynb} is given by
\begin{eqnarray}\label{dynb2}
p(x,t\,|\,y) = -k_1 \int_{0}^{t} p(x_1,s\,|\,y) G(x,t-s\,|\,x_1) \,ds+\,G(x,t\,|\,y).
\end{eqnarray}
The Laplace transform of equation (\ref{dynb2}) is given by
\begin{eqnarray}
\bar{p}(x,\tau\,|\,y) = -k_1 \bar{p}(x_1,\tau\,|\,y)
\bar{G}(x,\tau\,|\,x_1) \,ds+\,\bar{G}(x,\tau\,|\,y).
\end{eqnarray}
Thus,
\begin{eqnarray}
\bar{p}(x_1,\tau\,|\,y) = \frac{ \bar{G}(x_1,\tau \,|\,y)}{1+k_1
\bar{G}(x_1,\tau \,|\,x_1)}
\end{eqnarray}
and
\begin{eqnarray}
\bar{p}(x,\tau\,|\,y) = \frac{1}{k_1} +\frac{\bar{G}(x_1,\tau \,|\,y)-\bar{G}(x_1,\tau \,|\,x_1)-k^{-1}_{1}}{1+k_1
\bar{G}(x_1,\tau \,|\,x_1)}.
\end{eqnarray}
With
\begin{eqnarray}
\bar{G}(x,\tau \,|\,y) = \frac{1}{\tau}+\frac{2}{\pi}
\sum_{n=1}^\infty \frac{\cos nx\cos ny}{n^2 +\tau},
\end{eqnarray}
{ for $x,y \in (0,\pi)$, and to first order in $\tau$,  we
obtain}
\begin{eqnarray*}
\bar{p}(x,\tau\,|\,y)&=& \frac{1}{k_1} +\frac{\ds\frac{2}{\pi}
\sum_{n=1}^\infty \left[\ds\frac{\cos nx\cos ny}{n^2}
-\ds\frac{\cos^2 nx}{n^2}\right]
-k^{-1}_1}{k_1+\ds\frac{2k_1}{\pi} \sum_{n=1}^\infty \ds\frac{\cos
nx\cos ny}{n^2}}\tau +o(\tau)\\
&&\\
&=&\frac{1}{k_1} +\frac{\ds\frac{1}{2\pi}(x^2-y^2)
-k^{-1}_1}{k_1\left(1+\ds\frac{2}{\pi}\left(\ds\frac{\pi^2}{6}
-\ds\frac{\pi}{2}x +\frac{x^2}{2}\right) \right)}\tau+o(\tau).
\end{eqnarray*}
Hence
\begin{eqnarray*}
\bar{S}(\tau)& =& \frac{1-k_1\bar{p}(x_1,\tau \,|\,y)}{\tau}\\
&&\\
&=& \frac{\ds\frac{-1}{2\pi}\left(x^2-y^2\right)
+k^{-1}_1}{\left(1+ \ds\frac{2}{\pi}\left(\ds\frac{\pi^2}{6}
-\ds\frac{\pi}{2}x +\ds\frac{x^2}{2}\right) \right)}+o(1).
\end{eqnarray*}

Using the normalization condition that $S(0)=1$, {we find the
long time asymptotics}
\begin{eqnarray}
S(t) \sim_{r\rightarrow +\infty} \exp(-\alpha t),
\end{eqnarray}
where
\begin{eqnarray}\label{alphap}
\alpha^{-1} = \frac{ \ds\frac{-1}{2\pi}\left(x^2-y^2\right)
+k^{-1}_1}{\left(1+\ds\frac{2}{\pi}\left(\ds\frac{\pi^2}{6}
-\ds\frac{\pi}{2}x +\ds\frac{x^2}{2}\right)\right)}.
\end{eqnarray}
The un-normalized time constant for $x,y \in [0,L]$ is 
\begin{eqnarray}\label{alpha}
\alpha^{-1} = \frac{ \ds\frac{-\pi^3}{2D}\left(x^2-y^2\right)
+k^{-1}_1}{\left(1+\ds 2\pi\left(\ds\frac{1}{6}
-\ds\frac{1}{2L}x +\ds\frac{x^2}{L^2}\right)\right)}.
\end{eqnarray}
\subsubsection{The steady state limit}
{In the steady state, the mean number of molecules $N(t)$
surviving in $\Omega$ at time $t$ is}
\begin{eqnarray}
\lim_{t\rightarrow \infty} EN(t) = \lim_{\tau\rightarrow 0} \tau \bar{n}(\tau) = \gamma  \bar{S}(0)
\end{eqnarray}
Note that
 \beqq
\bar{S}(0) =\int_{0}^\infty \int_0^L  p(x,t|0)\, dx.
 \eeqq
If at time $t=0$ the particle is injected randomly at a point $y$,
we index $S$ by $y$ and { write}
 \beqq
\bar{S_{y}}(0) =\int_{0}^\infty \int_0^L  p(x,t|y) dx = \int_0^L G(x|y) dy
 \eeqq
{ where} $G(x\,|\,y)$ is { the} solution of
\begin{eqnarray*}
-\delta(x-y) &=& D\frac{\p^2 G}{\p x^2} -k \delta(x-x_1) G\\
&&\\
 \frac{\p G(0\,|\,y)}{\p x} &=& \frac{\p G(L\,|\,y)}{\p x} = 0.
\end{eqnarray*}
Thus,
\begin{eqnarray*}
G(x\,|\,y)= -\frac{\theta(x-y)}{D} +\frac{\theta(x-x_1)}{D}
+\frac{1}{k_1},
\end{eqnarray*}
{ where $\theta(x)$ is the integral of the Heaviside unit step function. It
follows that}
\begin{eqnarray*}
\bar{S}_y(0)=\int_0^L G(x\,|\,y) dy =  -\frac{(L-y)^2}{2D}
+\frac{(L-x_1)^2}{2D} +\frac{L}{k_1}.
\end{eqnarray*}
At steady state, the mean number of molecules surviving in
$\Omega$ at time $t$ is given by
\begin{eqnarray}
N_y(\infty)= \gamma \bar{S}_y(0)= \gamma (-\frac{(L-y)^2}{2D} +\frac{(L-x_1)^2}{2D} +\frac{L}{k_1})
\end{eqnarray}
{ If the particle is initially uniformly distributed}, $\Pr\{ y
\in[x,x+dx]\}=\ds\frac{dx}{L}$, then the steady state { mean
number of surviving molecules} is given by
\begin{eqnarray}\label{mean}
N(\infty)=\gamma E_y[\bar{S_{y}}(0)]=\gamma \left( -\frac{L^2}{6D}
+\frac{(L-x_1)^2}{2D} +\frac{L}{k_1} \right).
\end{eqnarray}

\subsubsection{Variance}
The second moment of the total { number of surviving particles
is defined as}
\begin{eqnarray*}
EN^2(t)=\sum_{n=1}^\infty n^2\Pr\{N(t)=n\}.
\end{eqnarray*}
We will now compute such number using the renewal equations. We have
\begin{eqnarray*}
EN^2(t)&=&\Pr\{x(t)\in \Omega \}\int_0^t f(s) k^2
\Pr\{N(t-s)= k-1\}\,ds \\
&&\\
&&+\Pr\{x(t)\notin [0,L] \}\int_0^t f(s) k^2
\Pr\{N(t-s)=k\}\,ds+\Pr\{x(t)\notin  \Omega\}\int_t^{\infty}
f(s)\,ds,
\end{eqnarray*}
which leads to
\begin{eqnarray*}
EN^2(t)&=&\Pr\{x(t)\in [0,L] \}\int_t^{\infty}
f(s)\,ds+\Pr\{x(t)\in \Omega \}\int_0^t EN^2(t-s) f(s)\,ds +\\ &
&\Pr\{x(t)\notin [0,L] \} \int_0^t EN^2(t-s) f(s)\,ds \\&+&
\sum_k\Pr\{x(t)\in \Omega \} \int_0^t(2k+1)Pr\{ N(t)=k-1 \}
f(s)\,ds.
\end{eqnarray*}
Thus,
\begin{eqnarray*}
EN^2(t) & = & \int_0^t EN^2(t-s) f(s)\,ds +\Pr\{x(t)\in \Omega \}\int_t^{\infty} f(s)\,ds+\\
& & 2\Pr\{x(t)\in \Omega\}\int_0^t E(N(t-s) f(s)\,ds +\Pr\{x(t)\in
\Omega\} \int_0^t f(s)\,ds
\end{eqnarray*}
and
\begin{eqnarray} \label{fluct}
EN^2(t) =  \int_0^t EN^2(t-s)  f(s)\,ds + \Pr\{x(t)\in \Omega\} +
2\Pr\{x(t)\in \Omega \}\int_0^t f(s) E(N(t-s)) ds.
\end{eqnarray}

{\bf \noindent Remark.}\par  
\noindent The variance can also be written in the form
\begin{eqnarray*}
&&\sigma^2_N(t)=EN^2(t)- [E(N(t))]^2 = \Pr\{x(t)\in  \Omega\} - \Pr\{x(t)\in  \Omega\}^2 +\\
&&\\
&&\int_0^t EN^2(t-s)  f(s)\,ds -\left( \int_0^t E(N(t-s)) f(s)\,ds
\right)^2.
\end{eqnarray*}

\subsubsection{The steady state variance}
To compute the variance in the steady state limit, as $t
\rightarrow \infty$, { we} Laplace transform equation
(\ref{fluct}),
\begin{eqnarray*}
\bar{n^2}(\tau) = \frac{\bar{S}(\tau)}{1-\bar{f}(\tau)}
+\frac{2}{1-\bar{f}(\tau)} \int^\infty_{0} \left[S(t)\int_{0}^t
f(t-s)n(s)\,ds\right] e^{-t\tau}\,dt.
\end{eqnarray*}
{It follows that}
\begin{eqnarray*}
EN^2(\infty)=\lim_{t\rightarrow \infty} EN^2(t)=  \lim_{\tau
\rightarrow 0}\tau\bar{n^2}(\tau).
\end{eqnarray*}
{ Writing}
\begin{eqnarray*}
\tau\bar{n^2}(\tau)= I(\tau) +II(\tau),
\end{eqnarray*}
{ we obtain the short $\tau$ approximation}
\begin{eqnarray*}
I(\tau)= \tau \frac{\bar{S}(\tau)}{1-\bar{f}(\tau)} \sim \gamma
\bar{S}(0).
\end{eqnarray*}
To { evaluate}
\begin{eqnarray*}
II(\tau)=\tau\frac{2}{1-\bar{f}(\tau)} \int^\infty_{0}
\left[S(t)\int_{0}^t f(t-s)n(s)\,ds\right] e^{-t\tau}\,dt,
\end{eqnarray*}
{ we recall} that $f(t) = \gamma e^{-\gamma t}$ and {
approximating $S(t)$ by $e^{-\alpha t}$ and interchanging the
order of integration, we obtain}
\begin{eqnarray*}
II(\tau)&=&\tau\gamma \frac{2}{1-\bar{f}(\tau)} \int^\infty_{0} n(s)e^{\gamma s}
\int_{s}^\infty  e^{-t(\gamma+\tau+\alpha )}dt, \\
&&\\
&=&\tau\gamma \frac{2}{1-\bar{f}(\tau)}\frac{1}{\gamma+\tau+\alpha } \int^\infty_{0} n(s)
e^{-t(\tau+\alpha )}dt,\\
&&\\
&=&\tau\gamma \frac{2}{1-\bar{f}(\tau)}\frac{1}{\gamma+\tau+\alpha }\bar{n}(\tau+\alpha )\\
&&\\
 &\sim& 2 \gamma^2\frac{1}{\gamma+\alpha }\bar{n}(\alpha ).
\end{eqnarray*}
Finally,
\begin{eqnarray*}
EN^2(\infty)=\gamma \bar{S}(0)+ 2 \gamma^2\frac{1}{\gamma+\alpha
}\bar{n}(\alpha )
\end{eqnarray*}
and
\begin{eqnarray} \label{noise}
\sigma^2 =EN^2(\infty)- \left(EN(\infty)\right)^2=\gamma
\bar{S}(0)-\gamma^2 \bar{S}^2(0) + \frac{2 \gamma^2}{\gamma+\alpha
}\bar{n}(\alpha ),
\end{eqnarray}
where  $\bar{S}(0)= \ds\frac{N(\infty)}{\gamma}$, { the time}
$\alpha$ depends only on $k_1$, the diffusion constant and the
length $L$, as given in formula (\ref{alpha}), and $\bar{n}(\alpha
) =\ds \frac{\gamma+ \alpha}{\alpha} \bar{S}(\alpha)$, where
$\bar{S}(\alpha )$ is the Laplace transform of the survival
probability at time $\alpha$.

\subsection{Push-pull chemical reaction in the continuum limit}
When the number of reacting molecules is large enough, regulated
by a  push-pull mechanism of hydrolysis and synthesis, the
reaction-diffusion equations is sufficient to  describe the
evolution of such system.  We assume that the molecules diffuse
inside a domain $\Omega$, but the synthesis occurs only uniformly
in a domain $\Omega' \subset \Omega$ while the hydrolysis is
performed by isolated enzymes. The hydrolysis can be modelled as a
killing measure $k(\x)$ (see \cite{HMS}). When there are a
discreet number $N_h$ of killing sources, the killing measure
$k(\x,t)$ is the sum of $\delta$ measures given by \beqq k(\x) =
\sum_{1}^{N_h} k_1\delta(\x-\x_k), \eeqq where $k_1$ is the
arrival rate constant to the killing sources. If the boundary of
the domain is divided into an absorbing part $\p\Omega_a$ and a
reflective part $\p\Omega_b$, the concentration $c(\x,t)$ can be
defined as the probability density function $p(\x,t)$ of the
system of N particles as \beqq c(\x,t)= Np(\x,t), \eeqq and \beqq
\frac{\p c(\x,t)}{\p t} &=& D \Delta c(\x,t) -k(\x,t) c(\x,t) +\gamma \chi_{\Omega'} \\
&&\\
\frac{\p c(\x,t)}{\p n}\bigg|_{\p \Omega_r }  &=& 0 ,\\
&&\\
c(\x,t)\bigg|_{\p \Omega_a}&=& 0, \\
&&\\
c(x,0) &=& c_0(x),
 \eeqq
where
 \beqq \int_{\Omega} c_0(\x) d\x = N.
  \eeqq
The mean number of particles $N(t)$ at any moment of time is given
by
 \beqq N(t) = \int_{\Omega} c(\x,t) dx.
  \eeqq
When the particles are contained inside a fixed domain and cannot
escape,$\Omega_a $ is empty. { The steady state density
$c(\x)=\lim_{t\to \infty}c(x,t)$ is a solution of}
 \beqq
0 &=& D \Delta c(\x) -k(\x) c(\x) +\gamma \chi_{\Omega'} \\
&&\\
0 &=& \frac{\p c(\x)}{\p n}\bigg|_{\p \Omega_r }.
 \eeqq
{ In one dimension} $\Omega'=\Omega=[0,L]$ and the killing
measure is reduced to a single point. The steady state equation is
\beq\label{sst}
0 &=& D\frac{\p^2 c(x)}{\p x^2} -k_1 \delta(x-x_1) c(x) +\gamma,\\
&&\nonumber\\
0 &=&\frac{\p c(x)}{\p x}\bigg|_{x=0} =\frac{\p c(x)}{\p
x}\bigg|_{x=L}, \nonumber
 \eeq
{ where} $k_1$ is the forward rate for one active site ({ it
has units of length per time}),  $\gamma$ is the uniform injection
rate of particle due to synthesis ({ it is the number of
injected particles per second per unit length}). Two integrations
of equation (\ref{sst}) lead to the solution
 \beq
c(x) = \gamma \left[\frac{L\theta(x-x_1)}{D} -\frac{x^2}{2D}
+\left(\frac{L}{k_1} +\frac{x_1^2}{2D}\right)\right],
 \eeq
where $\theta$ is the integral of the Heaviside function:
\begin{eqnarray*}
\theta(x)=\left\{
\begin{array}{lll}
0&&\hbox{for $x \leq 0$} \\
&&\\
x& &\hbox{for $x >0$}
\end{array}
\right.,
\end{eqnarray*}
Finally, the total number of free molecules is given by
\begin{eqnarray*}
N_0 =\int_{0}^L c(x)dx =\frac{\gamma L}{2D}(L-x_1)^2 -
\frac{\gamma L^3}{6D} +\gamma L \left(\frac{L}{k}
+\frac{x^2_1}{2D}\right),
\end{eqnarray*}
which should be compare to the steady state mean number obtained
in the discrete case (\ref{mean}).
\section{Markovian model of cell noise}\label{Markov}
We propose in this section an alternative model for computing the fluctuation of the number of bound molecules, based on a Markovian approach. We consider a domain $\Omega\subset\rR^3$ with $M$ mobile agonist
molecules and $S$ receptors, embedded in the boundary $\p\Omega$,
that open channels when they bind a single mobile agonist
molecule. We assume that the receptors occupy a small portion of
the surface area of $\p\Omega$. The agonists diffuse in $\Omega$
independently of each other. Bound agonists are released
independently of each other at exponential waiting times with rate
$k_{-1}$.

For a single receptor and a single agonist the time to binding is
the first passage time {to diffuse} to a small portion of the
boundary, $\p\Omega_a$, which is absorbing and represents the
active surface of the receptor, whereas the remaining part of
$\p\Omega$ is reflecting. It can be shown \cite{HS} that the
probability distribution of the first passage time to $\p\Omega_a$
is approximately exponential with rate
 \beqq
 \lambda_1=\frac{1}{\langle\tau_1\rangle},
 \eeqq
where $\langle\tau_1\rangle$ is the mean first passage time to
$\p\Omega_a$.

When there are $S$ channels, $k(t)$ of which are free at time $t$,
and $M$ agonists (gating molecules),
 \beq
 N(t)=(M-S+k(t))^+\label{Nk}
 \eeq
of which are free to diffuse in $\Omega$ at time $t$, where
\beqq
 x^+=\max\{0,x\}.
 \eeqq
Thus $$(S-M)^+\leq k(t)\leq S.$$

We assume that the pdf of the time for the next receptor to bind
is well approximated by the exponential pdf with instantaneous
rate
 \beqq
 \lambda_k(t)=\frac{N(t)k(t)}{\langle\tau_1\rangle}.
 \eeqq
This assumption is justified if the total area of the absorbing
boundary (the channels) is small relative to the surface area of
the {reflecting} boundary \cite{HS}.

It follows that when $N(t)=N$ and $k(t)=k$, the mean time to bind
is
 \beqq
\lambda_k=\frac{Nk}{\langle\tau_1\rangle}=\frac{k(M-S+k)^+}{\langle\tau_1\rangle}.
 \eeqq
The number of bound receptors at time $t$ is a birth-death process
with states $0,1,2,\dots,\min\{M,S\}$ and transition rates
 \beqq
 \lambda_{k\to k+1}=\lambda_k,\quad\lambda_{k\to
 k-1}=\mu=k_{-1}.
 \eeqq
The boundary conditions are
 \beqq
 \lambda_{S\to S+1}=0,\quad \lambda_{0\to-1}=0.
 \eeqq

Setting
 \beqq
 P_k(t)=\Pr\{k(t)=k\},
 \eeqq
we have  \cite{Saaty}
 \beqq
\dot P_{(S-M)^+}(t)&=&-k_{-1}SP_{(S-M)^+}(t)+\lambda_{1}P_{(S-M)^++1}(t) \\
&&\\
 \dot
P_k(t)&=&-\left[\lambda_{k}+k_{-1}(S-k)\right]P_k(t)+\lambda_{k+1}P_{k+1}(t)+k_{-1}(S-k+1)P_{k-1}(t)\\
&&\\
&&\quad\mbox{for}\quad
k=(S-M)^++1,\dots,S-1\nonumber\\
&&\nonumber\\
\dot P_S(t)&=&-\lambda_SP_S(t)+k_{-1}P_{S-1}(t).
 \eeqq
The initial condition is $P_{k,q}(0) = \delta_{k,S} \delta_{q,0}$.
In the steady state the average number of open channels is
 \beqq
 \langle k_\infty\rangle=\sum_{j=(S-M)^+}^{S}jP_j,
 \eeqq
where $P_j=\ds\lim_{t\to\infty}P_j(t)$, and the stationary
variance in the number of open channels is determined from the
second moment
 \beqq
 \langle k^2_\infty\rangle=\sum_{j=(S-M)^+}^{S}j^2P_j
 \eeqq
by
 \beqq
 \sigma^2(M,S)=\langle k_\infty^2\rangle-\langle k_\infty\rangle^2.
 \eeqq

\subsection{The steady state approximation}
In the steady state
    \beqq
0&=&-k_{-1}SP_{(S-M)^+}+\lambda_{1}P_{(S-M)^++1} \\
&&\\
0&=&-\left[\lambda_{k}+k_{-1}(S-k)\right]P_k+\lambda_{k+1}P_{k+1}+k_{-1}(S-k+1)P_{k-1}\\
&&\\
&&\quad\mbox{for}\quad
k=(S-M)^++1,\dots,S-1\nonumber\\
&&\nonumber\\
0&=&-\lambda_SP_S+k_{-1}P_{S-1},
 \eeqq
which gives for $(S-M)^+\leq k\leq S$
 \beqq
 P_{S-1}&=&P_S\frac{\lambda_S}{k_{-1}}\\
 &&\\
 P_{S-k}&=&P_S\frac{\prod_{i=S-k+1}^S\lambda_i}{k!k_{-1}^k}=P_S\frac{\prod_{i=S-k+1}^S i(M-S+i)^+}{k!(\langle\tau_1\rangle k_{-1})^k}\\
 &&\\
 P_{(S-M)^+}&=&P_S\frac{\prod_{i=(S-M)^++1}^S\lambda_i}{(S-(S-M)^+)!k_{-1}^{S-(S-M)^+}}
 =P_S\frac{\prod_{i=(S-M)^++1}^S i(M-S+i)^+}{(S-(S-M)^+)!(k_{-1}\langle \tau_1
 \rangle)^k}.
 \eeqq
{ The} constant $P_0$ is determined from the normalization
condition
 \beqq
 \sum_{k=(S-M)^+}^{S}P_k= \sum_{k=S-(S-M)^+}^{0}P_{S-k}    =1.
 \eeqq
Thus,
 \beq \label{moments}
 P_S &=& \frac{1}{1+\sum_{k=1}^{S-(S-M)^+} \ds\frac{\prod_{i=S-k+1}^S \lambda_i}{k!k_{-1}^k}}=\ds
\frac{1}{1+\ds\sum^{S-(S-M)^+}_{k=1}\ds\frac{\prod_{i=S-k+1}^S
i(M-S+i)^+ }{k!(\langle\tau_1\rangle k_{-1})^k }}
\nonumber\\
& & \nonumber \\
\langle k_\infty \rangle &=&   \sum_{k=S-1}^{(S-M)^+} (S-k)^+ P_{S-k} =  P_S  \sum_{k=S-1}^{(S-M)^+} (S-k)^+ \frac{\prod_{i=S-k+1}^S i(M-S+i)^+}{k!(\langle\tau_1\rangle k_{-1})^k}
\nonumber  \\
& &\nonumber\\
 \langle k^2_\infty\rangle&=&  \sum_{k=S-1}^{(S-M)^+} [(S-k)^+]^2 P_{S-k} =  P_S  \sum_{k=S-1}^{(S-M)^+} [(S-k)^+]^2 \frac{\prod_{i=S-k+1}^S i(M-S+i)^+}{k!(\langle\tau_1\rangle k_{-1})^k} \nonumber \\
& &\nonumber\\
\sigma^2_S(M) &=&  \langle k^2_\infty\rangle - \langle
k_\infty\rangle^2.
 \eeq
The graph of $\sigma^2_S(M)$ {\em vs} $M$ is given  in Figure
\ref{varianceMarkov}.

\subsection{The steady state mean number of open channels}
Using the above Markov model an explicit formula can be
obtained for the mean number of open channels ( that bind an
agonist molecule). It is assumed here that a channel can bind only
a single agonist molecule at a time. This approach does not give
any information about the fluctuations.

{ The steady state fraction $n^*_1$ of bound agonist molecules
in a bounded domain $\Omega$ that contains $S$ channels and $M$
gating agonists is}
 \beq
n^*_1 & =&\frac{k_1}{k_1+k_{-1}} M,\quad \\
& & \label{equil}
\\
n_1   & =&\frac{k_{-1}}{k_1+k_{-1}}M,
 \eeq
where $k_1$ is the forward binding rate in solution and $k_{-1}$
is the backward rate. If $S_{ch}$ is the total surface occupied by
the channels, then $S_{ch} = S S^1_{ch}$, where $S^1_{ch}$ is the
effective surface occupied by a single channel.  $S^1_{ch}$ is
defined by the area where the electrostatic force of the channel
is { sufficient to bind an agonist}. Recall that $k_1$ is by
definition { the mean} number of { agonist molecules}
arriving at the channel per unit of time. Using the result of the
previous section, { we can write}
 \beq\label{iden} k_1
=\frac{S_{ch}}{\tau_{n_1 }(M-S+n_1)^+}.
 \eeq
Thus, {using equations (\ref{equil}) and (\ref{iden}), we find
that the mean number of bound agonist, which is the same as that
of bound channels}, is given by
 \beq
n^*_1&=&\frac{1}{1+\ds\frac{\tau_1
k_{-1}}{S_{ch}}(M-n^*_1)(S-n^*_1)}M,
 \eeq
{ with
 \beq
 \langle \tau_1 \rangle =\frac{|\Omega|}{\pi D} \log
\frac{S_{ch}}{|\p \Omega|},
 \eeq
where $|\Omega|$ is the volume of the $\Omega$} and $|\p \Omega|$
is the surface area { of its boundary} (see \cite{HS}). For
$n^*_1$ small relative to $M$ and $S$,
 \beq
n^*_1 & \sim &\frac{1}{1+\ds\frac{\tau_1 k_{-1}}{S_{ch}}MS}M\\
&&\\
& = &\frac{1}{1+\ds\frac{\tau_1 k_{-1}}{S^1_{ch}}M}M,
 \eeq
which gives explicitly the number of bound channels as a function
of the geometrical parameters. A similar formula can be derived,
when a single channel can be bound to several gating molecules.

\subsection{The Michaelis-Menten law in micro-structures}
The rate of production of a product $P$ from a substrate $M$ by a
catalytic enzyme $E$ is usually described  in text books by the Michaelis-Menten
law for reactions in solution. In confined micro-domains, the above analysis can be used to estimate the
number of $P$ produced in the reaction.

The  chemical reaction is described by
\begin{equation}
M+E
\begin{array}{l}
k_{f1} \\
\rightleftharpoons \\
k_{b1}%
\end{array}%
ME
\begin{array}{l}
\overset{k_{b2}}{\rightharpoonup } \\
\end{array}
E+P.  \label{MMR2}
\end{equation}
and a master equation for the joint probability that the number of
$P$ molecules produced is $k$ and $q$ enzymes are bound { at
time $t$}, $ P_{k,q}(t) = Pr \{P(t)=k, E(t)=q \}$, can be derived
{ as above}. In particular, {the kinetics of the reaction
(\ref{MMR2}) are}
 \beqq  \label{MM1}
\dot{P}_{k,q}(t) &=& (1-q(k_{-1}+k_2)-\lambda_{M,q} )P_{k,q}(t) +
k_2 (q+1)
P_{k-1,q+1}(t) +k_{-1}(q+1)P_{k,q+1}(t)\\
&&\\
&&+ \lambda_{M,q-1} P_{k,q-1}(t),\quad\hbox{ for }  1\leq q, k \leq E_0-1 \\
&&\\
\dot{P}_{0,q}(t) &=& (1-\lambda_{M,q} )P_{0,q}(t) +k_{-1}(q+1)P_{0,q+1}(t)+
\lambda_{M,q-1} P_{0,q-1}(t),\\
&&\\
\dot{P}_{k,E_0}(t) &=& (1-k(k_{-1}+k_2) )P_{k,E_0}(t) + k_2 q P_{k-1,E_0}(t)
+ \lambda_{M,E_0-1} P_{k,q-1}(t),\\
&&\\
\dot{P}_{M_0,0}(t) &=& (1-M_0(k_{-1}+k_2) )P_{M_0,0}(t) + k_2
P_{M_0-1,0}(t),
  \eeqq
{with the initial condition ${P}_{k,q}(0) = \delta_{M-0,0}$
and}
 \beqq
\lambda_{M,q} &=& \frac{M(E_0-k)}{\tau_1},\\
&&\\
M &=& (M_0-k-q)^+.
 \eeqq
{Here} $M_0$ is the initial number of substrate {
molecules}, $M$ is the number of { unbound} substrate
molecules, and $E_0$ is the total number of enzyme available. The
mean  number of $P$ produced at time $t$ is
 \beq
\langle P(t) \rangle = \sum_k k Pr\{ P(t)=k \} =  \sum_k k \sum_q Pr\{ P(t)=k,E(t)=q \}.
 \eeq
The only steady state solution of equation (\ref{MM1}) is zero as any steady state.

\subsection{Fluctuations in a push-pull system with binding}
We consider now a push-pull system, where gating molecules can
also bind and unbind to some proteins. {We consider two main
approaches to the description of this dynamics}. The first
approach is based on the procedure used in the first model in {Section} \ref{firstM} 
and consists in deriving an equation for the
joint probability of a trajectory, the number of bound sites in
the volume $\Delta \x$, and the total number of { molecules} in
$\Omega$,
 \beq
p(\x,S,M,t\,|\,\y)\,\Delta\x= \Pr\left\{\x(t)\in\x+\Delta\x,\,
S_{\Delta}(\x,t)=S\,, M |\,\x(0)=\y\right\}.
\label{jpdfp}
 \eeq
The function $p(\x,S,M,t\,|\,\y)$ is the joint probability density
to find an agonist and $S$ free binding sites at $\x+\Delta \x$ at
time $t$, and $M$ agonists in the domain, conditioned by the
initial position $\y$ of the agonist. An analysis { similar to
that of Section  \ref{firstM} leads to}
 \beq \frac{\p p(\x,S,M,t)
}{\p t} &=& -\nabla\cdot\mb{J}(\x,S,M,t)
 - K_1 Mp^2(\x,S,M,t)S -k_{-1}[S_0(\x)-S]p(\x,S,M,t) \nonumber\\
&&\nonumber\\
&+& K_1 M
(S+1)p^2(\x,S+1,M,t)+k_{-1}[S_0(\x)-S+1]p(\x,S-1,M,t)+\nonumber\\
&&\nonumber\\
 & &\gamma p(\x,S,M,t) -(M-S)^+
K_{-1}p(\x,S,M,t)\label{eqfff},
 \eeq
where by definition $\mb{J}(\x,S,M,t)$ is the joint probability flux
at position $\x$ at time $t$, and $S$ proteins are free for $M$ agonists. It is
defined in the diffusion case by
 \beq \mb{J}(\x,S,M,t) = -D \nabla p(\x,S,M,t).
 \eeq
{The} new (forward) binding rate is
\begin{eqnarray*}
K_1 = k_1 \Delta\x,
\end{eqnarray*}
$\gamma$ is the uniform production rate, and $K_{-1}$ is the
uniform killing rate for the $(M-S)^+$ free {agonist
molecules. The moments can be calculated from $p(\x,S,M,t)$, as
above.}

The second procedure consists in adding directly the push-pull
effect in the Markovian model. Defining the joint probability that
$k$ channels are free and $M(t)$ are in the reacting domain at
time $t$,
 \beq P_{k,q}(t)=\Pr\{k(t)=k,M(t)=q \},
  \eeq
{ neglecting} the distribution of the sources and sinks
responsible for the fluctuation of the number of molecules, {
we consider} only the case { that} hydrolysis and synthesis
{ occur at exponential waiting times (i.e., are Poissonian),
with rates $K_{-1}$ and $\gamma$, respectively}. Following the
same steps as in Section \ref{Markov}, the master equation for
$P_{k,q}(t)$ becomes
 \beqq
\dot P_{(S-M)^+,q}(t)&=&-k_{-1}SP_{(S-M)^+,q}(t)+\lambda_{1,q}(q-S+1)^{+}P_{(S-M)^++1,q}(t) \\
&&\\
\dot P_{k,q}(t)&=&-(\lambda_{k,q}+(S-k)k_{-1} +\gamma+K_{-1}q)P_{k,q}(t)+\\&&\\
& &\lambda_{k+1,q}P_{k+1}(t))+k_{-1}(S-k+1)P_{k-1,q}(t)+ \gamma P_{k,q-1} +K_{-1}(q+1)P_{k,q+1},\\&&\\
&&\mbox{for}\quad k=(S-M)^++1,\dots,S-1,\nonumber
 \eeqq
where we recall that
\beqq
\lambda_{k,q} =\frac{Nk}{\langle\tau_1\rangle}=\frac{k(M-q+k)^+}{\langle\tau_1\rangle}.
 \eeqq
The initial condition is $P_{k,q}(0) = \delta_{k,S} \delta_{q,0}$.

Finally, if the push-pull rate is much slower than the binding
rate, the previous results of the Markovian model can { be}
used directly and the mean number of free channels, $N_\infty(t)$,
is directly computed using the moments (\ref{moments}) and
 \beqq
N_\infty(t) = \sum_{M} \langle k_\infty\rangle(M) \Pr\{N(t)=M \},
\eeqq where $\langle k_\infty\rangle(M)$ is the mean number of
bound channels when there are $M$ gating molecules in { the
domain,}
 \beqq
\langle k_\infty\rangle(M) = \sum_{k=S-1}^{(S-M)^+} (S-k)^+
P_{S-k}(q).
 \eeqq
{ Here, by definition, the dependence of $P_{S-k}$ on $q$ is
that obtained} in equation (\ref{moments}), and $\Pr\{N(t)=m \}$
is the probability that $M$ molecules are in the reacting domain
at time $t$ (see Section \ref{ppm}). In the limit $t\to\infty$
 \beqq
\langle N_\infty(\infty)\rangle &=& \sum_{M} \langle k_\infty\rangle(M) \Pr\{N(\infty)=M \}\\
&&\\
 \langle N^2_\infty(\infty)\rangle &=& \sum_{M} \langle
k_\infty\rangle^2(M) \Pr\{N(\infty)=M \}. \eeqq

\section{Conclusion and biological implications}

\subsection{Comparison of the models}
We presented here two models that describes fluctuation due to
binding and binding of agonist to some fixed  proteins.  In the
Markovian model, the only { geometric feature of the cell that
enters the model is the cell's volume}. { The distribution of
the channels, as well as other geometric features are ignored}. The
{ advantage} of such approach is that it gives explicit
estimates of the mean and the variance (see equation \ref{moments}). In the first model, more
details of the geometry and organization of the channel { are captured, however, the computation of the moments requires the solution of a} system of partial differential equations. The expression of the variance 
is given in general by equation \ref{sd} and in steady state by expression \ref{Var}.

\subsection{The Forward binding rate}
The Markovian model does not rely on the forward binding rate and
it is replaced here by the mean  arrival time to the channels
$\langle\tau_1\rangle$. The initial forward binding { rate} per
molecule is the arrival rate of a gating molecule { to any one
of the binding sites on the $N$ proteins}, and can be defined as
 \beq K_1 =
\frac{1}{\langle\tau_1\rangle N}.
 \eeq
{ The number $N$ can be, for example, the number of proteins in
a given volume. It} can be computed from the concentration $[N]$.
In that case $K_1$ can be rewritten as
 \beq
K'_1 = \frac{1}{\langle\tau_1\rangle[N]}.
 \eeq
The Markovian approach proves that the traditional forward binding
rate has to be { abandoned} and a dynamical rate has to be used
{ instead}. The first model uses another definition of the
forward binding rate, that can be related to the traditional one.

\subsection{Biological implications}
The mean and the variance of the number
of bound molecules were derived in the first model for a { finite system of
molecules.} In particular formula (\ref{sigma2}) proves that the
variance depends { on the total number of bound molecules at
time} $t$ $\bar{S_0}(t)$ ({ see} formula (\ref{Pro})). {
Note} that $\bar{S_0}(t)$ is the integral of the joint probability
{ density function} $p(x,1,t)$, which is a solution of a
partial differential equation. Thus the number of bound molecules
is a complicated function of the geometry, the distribution of the
substrate molecule and the interaction between the molecules
through the number of binding sites.

As seen in the experimental data \cite{Yau,Koko,matthews,sigworth}, the variance
(\ref{sigma2}) does not only reflect binding to proteins, but {
also} a dynamical process of binding and unbinding in a
microstructure, which involves the geometry, the distribution of
the protein and the diffusion process. Only at high concentration
of binding molecules, a molecule that leaves a binding site will be
immediately replaced by another gating molecule. In the regime
{ of high concentration}, when the protein represents a gating
channel located on the { cell} membrane, the fluctuation of the
current represents { an} intrinsic property of the channel.
The fluctuations, then, are proportional to the gating property of
the channel. But  for  many sensor cells, the concentration of
gating molecules is not high, the small number of gating molecules 
is the cause of perpetual fluctuations due to binding and unbinding.

\begin{figure}
\centering \psfig{figure = 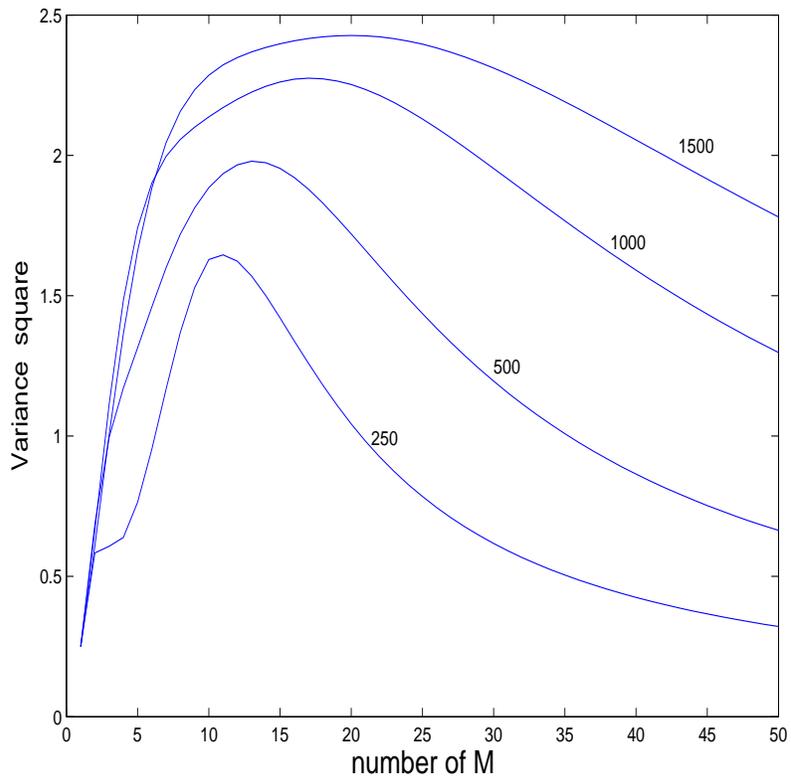, height=105mm,width=105mm} 
\caption{\small {The variance square equal to $\sigma^2_S(M)$,
for various values of the backward binding rate
$k_{-1}=[250,500,1000,1500]$. The total number of substrate
molecules is fixed at $S=10$ and $\tau=0.01ms$.} }
\label{varianceMarkov}
\end{figure}

\begin{figure}
\centering \psfig{figure = 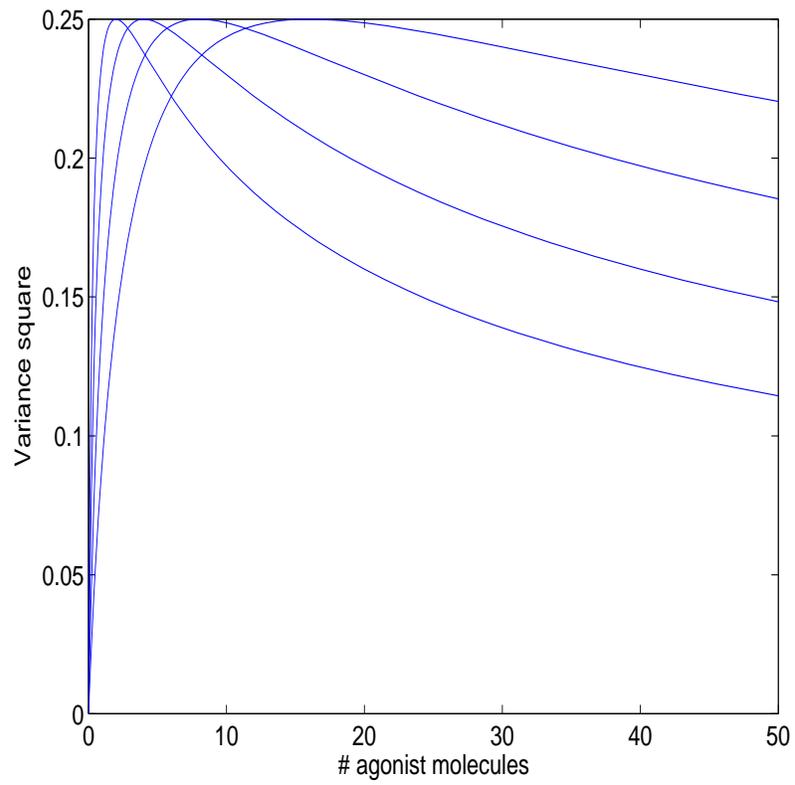, height =105mm,width=105mm} 
\caption{\small {Normalized variance $\sigma^2$, for $\frac{k_1 \Delta \x}{N_S k_{-1}}=$ 1/2, 1 ,2 ,4}}
\label{variance1}
\end{figure}

The graphs \ref{variance1} and \ref{varianceMarkov} of the current noise variance {\em vs} $\bar{S_0}(t)$
has almost the shape of an inverted parabola, which implies that there it has unique maximum point at a specific location. In the steady state limit, it is obtained in case of the first model for $\bar{S_0}(t)=1/2$. Biological microsystems, such as photo-receptors, usually operate at low noise levels, far from the maximum of the graph.

Controlling the membrane voltage fluctuation for sensor cells is a
crucial issue for the transduction process that consist in
transmitting a molecular signal. Channels fluctuation is not
responsible only for the noise of the cell, but the biochemistry
underneath and the organization of the cell play a crucial role.
In a set of experiments \cite{Yau,Koko,matthews,sigworth}, the
membrane noise was measured but the purpose was to identify the
properties of the channel rather than the molecular dynamics. Using
the results of the previous mathematical analysis, it is a hard
problem to convert the data of fluctuation obtained for a detached
patch experiment to the fluctuation inside a single cell. A
separate mathematical analysis has to be performed to identify
such fluctuation.

\begin{figure}
\centering \psfig{figure = 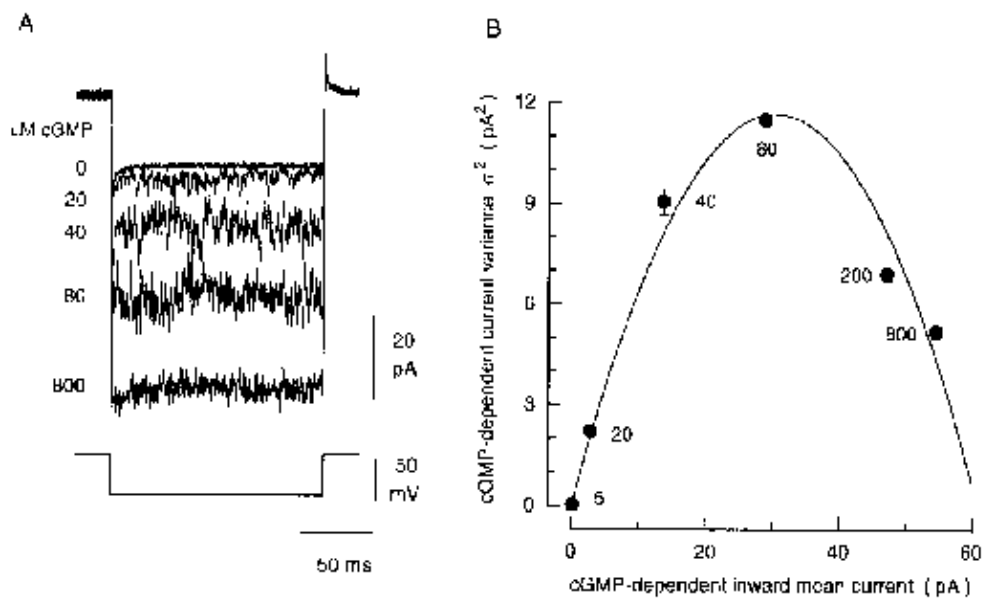}
\caption{{\small {{\bf Experimental analysis of the fluctuation in the number of open channel.} The left panel corresponds to the current response for various concentration of the gating molecules(cGMP), in a detached patch experiment of a cone Photoreceptors membrane. On the right panel, the variance is plotted versus the mean current, which is proportional to the number of gating molecules(cGMP). This picture was published in \cite{Koko} figure 2. }}}
\label{variance1Koko}
\end{figure}

In the experimental data, the variance of the fluctuation has been
related \cite{sigworth} to the total current, when the probability
of the channel to be in an open state is independent from the
other. When the gating molecule is not as abundant, channels share
such resource and when a gating molecule is bound to a channel,
this resource is not available for the neighboring channels. This
effect coupled the probability of the channel to be in open state
through the number of gating molecules. It is useful to compare the
 experimental result \ref{variance1Koko}, with the simulations of the theoretical model,
 obtained in figures  \ref{variance1}  and \ref{varianceMarkov}. This comparison  suggests 
that the tail distribution of the variance cannot be approximated everywhere by a parabola, 
as it  was done, under the assumption that the opening of the channels is independent. This 
assumption breaks at high  concentration, where the channels are coupled through 
the gating molecules.

In a system composed by a cell membrane containing channels and
gating molecules only, two time scales should be dominant. The
first time scale is related to the time it takes for a gating
molecule to find a channel. This time depends on the geometry of
the domain, the number of gating molecules and the number of
channels: for few molecules, this time can be approximated by the
mean time it takes for a gating molecule to find a small absorbing
boundary. This problem has been treated in \cite{HS} and this
approximation was used in the Markovian model of the paper to
estimate the number of open channels. This
assumption is valid because channels or binding proteins occupy a
small portion of the boundary surface. The second time scale is
related to the backward binding constant of the gating molecule to
the channel. The backward binding constant depends on the property
of the channel only and thus does not depend on the statistical
property of the system.

\noindent {\bf Acknowledgments:} We would like to thank J. Korenbrot and R. Nicoll for stimulating discussions. D. H thanks the Sloan-Swartz foundation for the financial support. Part of this work was done while he was visiting UCSF.


\newpage

\newpage

\newpage
\newpage

\newpage

\newpage

\end{document}